\documentclass{elsart}
\usepackage{graphicx,amssymb}
\journal{Physica B}
\begin{document}
\begin{frontmatter}
\title{Breakdown of an intermediate plateau in the magnetization process 
         of anisotropic spin-1 Heisenberg dimer: theory vs. experiment}
\author[label1]{J. Stre\v{c}ka\corauthref{cor1}}, 
\ead{jozef.strecka@upjs.sk, jozkos@pobox.sk}
\author[label2]{M. Hagiwara}, 
\author[label1]{P. Bal\'a\v{z}},
\author[label1]{M. Ja\v{s}\v{c}ur},
\author[label4]{Y. Narumi}, 
\author[label2]{S. Kimura}, 
\author[label3]{J. Kuch\'ar}, and
\author[label4]{K. Kindo}
\address[label1]{Department of Theoretical Physics and Astrophysics, 
Faculty of Science, \\ P. J. \v{S}af\'{a}rik University, Park Angelinum 9,
040 01 Ko\v{s}ice, Slovak Republic}
\address[label2]{KYOKUGEN (Center for Quantum Science and Technology under Extreme Conditions), 
Osaka University, 1-3 Machikaneyama, Toyonaka, Osaka 560-8531, Japan}
\address[label4]{Institute for Solid State Physics, University of Tokyo, \\ 
5-1-5 Kashiwa-no-ha, Kashiwa, Chiba 277-8581, Japan}
\address[label3]{Department of Inorganic Chemistry, Faculty of Science, \\ 
P. J. \v{S}af\'{a}rik University, Moyzesova 11, 041 54 Ko\v{s}ice, Slovak Republic}
\corauth[cor1]{Corresponding author.} 
             
\begin{abstract}
The magnetization process of the spin-1 Heisenberg dimer model with the uniaxial or biaxial single-ion anisotropy is particularly investigated in connection with recent experimental high-field 
measurements performed on the single-crystal sample of the homodinuclear nickel(II) compound [Ni$_2$(Medpt)$_2$($\mu$-ox)(H$_2$O)$_2$](ClO$_4$)$_2$.2H$_2$O (Medpt=methyl-bis(3-aminopropyl)amine). 
The results obtained from the exact numerical diagonalization indicate a striking magnetization 
process with a marked spatial dependence on the applied magnetic field for arbitrary but 
finite single-ion anisotropy. It is demonstrated that the field range, which corresponds 
to an intermediate magnetization plateau emerging at a half of the saturation magnetization, 
basically depends on a single-ion anisotropy strength as well as a spatial orientation of 
the applied field. The breakdown of the intermediate magnetization plateau is discussed 
at length in relation to the single-ion anisotropy strength. 
\end{abstract}

\end{frontmatter}

\section{Introduction}

Over the last few decades, the quantum behaviour of low-dimensional molecule-based magnetic materials 
has become one of the most fascinating fields emerging at the border of condensed matter physics, 
inorganic chemistry and materials science \cite{design}. A vigorous scientific interest aimed at 
the molecule-based magnetic materials arises due to their wide potential applicability. In particular, the molecule-based magnets might possibly serve as useful magnetic and optical devices intended 
for a molecular electronics (molecular switches and other useful magneto-optical devices) \cite{opto}, high-density storage devices designed for a computer science \cite{storage}, or basic entities 
that are suitable for a quantum computation \cite{qc}. One of the attractive features of 
molecule-based magnetic materials embodies a recent advance achieved in an attempt to take control 
over the connectivity and magnetic architecture of synthesized molecule-based materials, which are specifically tailored through judicious choice of appropriate molecular building blocks. 
This recent progress in a molecular engineering seems to be an essential ingredient by approaching 
a targeted design of magnetic materials with desired magnetic properties \cite{design}. 

Small magnetic spin clusters (SMSCs), which denote weakly interacting assemblies of molecules formed 
by a few exchange-coupled paramagnetic centres (spins), belong to the simplest molecule-based 
magnetic materials \cite{cluster}. Note that the discrete magnetic molecules are held together 
to form a molecular crystal merely by virtue of van der Waals forces and/or hydrogen bonding. Accordingly, SMSCs are of particular research interest since they often allow accurate 
description of their magneto-structural correlations with respect to utterly negligible inter-molecular interactions. In addition, SMSCs are of immense practical importance as they provide excellent 
testing ground for a deeper understanding of remarkable cooperative quantum phenomena. 

The main obscurities, which still remain unresolved in the area of SMSCs, are mostly closely associated with quantum manifestations of molecule-based materials with a pronounced magnetic anisotropy, i.e. molecular crystals with highly spatially dependent magnetic properties. In recent years, a considerable attention has been paid to single-molecule magnets \cite{smm} and single-chain magnets \cite{scm}. 
This scientific interest has been motivated by the idea of having denser information-recording media allowing data storage several orders of magnitude greater than at present \cite{storage}. As a rule, the recorded information is destructed through a competitive spin tunneling effect, since the antiferromagnetic state is thermodynamically more stable than the ferromagnetic one. The energy barrier must be therefore large enough to prevent such degradation of information and hence, the magnetic anisotropy should be sufficiently strong. Owing to this fact, the major scientific interest is currently focused on a fundamental understanding of the magnetic anisotropy and its role by 
determining the overall magnetic behaviour of SMSCs \cite{anis}.

Our previous analytical calculations have been  concerned with anisotropic properties of 
the spin-1 Heisenberg dimer, which can serve as a suitable model to a rich variety of existing 
homodinuclear nickel(II) complexes (see Ref. \cite{stre05} and references cited therein).
For comparative purposes, our theoretical predictions have been also confronted with recent experimental measurements performed on the single-crystal sample of the homodinuclear nickel(II) coordination compound [Ni$_2$(Medpt)$_2$($\mu$-ox)(H$_2$O)$_2$](ClO$_4$)$_2$.2H$_2$O (Medpt = methyl-bis(3-aminopropyl)amine) to be further abbreviated as NAOC. This complex has been chosen 
as a typical representative of the spin-1 dimeric compounds for at least two reasons. First, it exhibits highly anisotropic magnetic properties and secondly, the notable high-field magnetization 
data displaying entire two-step magnetization curve are available for this compound \cite{Narumi}. 
As could be expected for nickel-based coordination compounds such as NAOC complex, our previous 
study has revealed a relatively strong effect of uniaxial single-ion anisotropy and the utterly negligible influence of the exchange anisotropy. It has been actually shown that a striking 
magnetization process with a marked dependence on the spatial orientation of the applied field 
arises almost exclusively on account of the single-ion anisotropy effect, which is, on the other hand, too small to cause the breakdown of intermediate magnetization plateau theoretically predicted in Ref. \cite{stre05}. It is worthy of notice that the high-field ESR measurement performed on the single-crystal sample of NAOC has provided a strong indication of the non-negligible 
biaxial single-ion anisotropy \cite{ESR}. 

In the present article, we will employ the exact numerical diagonalization in order to clarify 
the magnetization process of the spin-1 Heisenberg dimer with the uniaxial single-ion anisotropy 
in the magnetic field oriented perpendicular to a quantization axis (i.e. by applying the transverse magnetic field), as well as, the magnetization process of the spin-1 Heisenberg dimer with the biaxial single-ion anisotropy for both parallel as well as perpendicular field directions. Notice that both investigated magnetization curves bring new insight into how the single-ion anisotropy determines the anisotropic properties of homodinuclear nickel complexes. It should be also pointed out that these characteristic features of the spin-1 dimer model are inaccessible within the exact analytical diagonalization, which has been used to obtain the magnetization curves of the spin-1 Heisenberg 
dimer with the uniaxial single-ion anisotropy in the longitudinal magnetic field oriented along a quantization axis \cite{stre05}. 
  
The organization of this paper is as follows. In Section \ref{model}, we will introduce the model system by defining its Hamiltonian and we also recall the foundations for the occurrence of the magnetic anisotropy. The most interesting numerical results obtained for the magnetization process under different spatial orientations of the applied external field are presented and detailed 
discussed in Section \ref{result}. In the next section, we provide a comparison between the 
obtained theoretical results and the relevant experimental measurements on NAOC compound. 
Finally, some concluding remarks are given in Section \ref{conclusion}.
  
\section{Model and its Hamiltonian}                                                                     
\label{model}

First, let us make few remarks about possible sources of the magnetic anisotropy that come into question in the spin-1 dimer model designed for describing magnetic features of the homodinuclear nickel(II) coordination compounds. Since the ground state of divalent nickel ion is in an 
octahedral environment orbitally non-degenerate, more or less isotropic intra-dimer interaction 
should be expected within the molecular entities containing dinickel cores \cite{anis}. It is quite obvious from the aforementioned argument that the main contribution to the overall magnetic 
anisotropy should come from the single-ion anisotropy effect \cite{zfs}. This finding is fully consistent with our previous study, which has demonstrated in NAOC compound a relatively strong uniaxial single-ion anisotropy and utterly negligible exchange anisotropy \cite{stre05}. 
It should be remembered, however, that the single-ion anisotropy comes from the low-symmetry 
crystal field of ligands creating a coordination sphere of the nickel centres and thus, 
there should be obvious structural indications (reflected in bond angles and bond lengths) 
if a rather high magnetic anisotropy is observed \cite{ma}.  

With all this in mind, the magnetic behaviour of the homodinuclear nickel(II) complexes will be interpreted with the aid of the isotropic spin-1 Heisenberg dimer model refined by the zero-field splitting parameters that account for the uniaxial or biaxial single-ion anisotropy. It is worthwhile to remark that there exist several exactly soluble limiting cases of this simple model system. 
In a presence of the external magnetic field applied along the quantization axis (longitudinal field), the complete exact analytical solution can be found both for the isotropic spin-1 Heisenberg dimer \cite{isotropic}, as well as, the spin-1 Heisenberg dimer with the uniaxial single-ion anisotropy \cite{isotropicd}. On the other hand, the effect of transverse magnetic field (i.e. the external field oriented perpendicular with respect to the quantization axis) has not been dealt with in the literature so far even for a such simple system as the spin-1 Heisenberg dimer with the uniaxial single-ion anisotropy. Therefore, the main goal of the present work is to explore the magnetization process 
in the transverse magnetic field for the spin-1 Heisenberg dimer with either uniaxial or biaxial single-ion anisotropy. 

Let us start by defining the following effective spin Hamiltonian
\begin{eqnarray}
\hat {\mathcal H} &=& 
   J (\hat S_1^x \hat S_2^x + \hat S_1^y \hat S_2^y + \hat S_1^z \hat S_2^z) 
   + D [(\hat S_1^z)^2 + (\hat S_2^z)^2] \nonumber \\
&+& E [(\hat S_1^x)^2 - (\hat S_1^y)^2] + E [(\hat S_2^x)^2 - (\hat S_2^y)^2] \nonumber \\
&+& g_x \mu_{\mathrm{B}} B_x (\hat S_1^x + \hat S_2^x)
 +  g_z \mu_{\mathrm{B}} B_z (\hat S_1^z + \hat S_2^z), 
\label{eq:ham}	   
\end{eqnarray}
where $\hat S_1^{\alpha}$ and $\hat S_2^{\alpha}$ ($\alpha$ = $x$, $y$, or $z$) denote the spatial components of the local spin-1 operator on the metal centres 1 and 2, $J$ is the isotropic Heisenberg exchange interaction between them and finally, the anisotropy constants $D$ and $E$ label the 
zero-field splitting parameters that account for the uniaxial and biaxial single-ion anisotropy, respectively. Other terms have an usual meaning: $B_\alpha$ is the external magnetic field applied along the $\alpha$-axis ($\alpha = x, z$), $\mu_{\mathrm{B}}$ stands for Bohr magneton and 
$g_{\alpha}$ denotes a spatial component of the $g$-factor. For easy reference, we will further 
refer to the $z$-axis as to the quantization axis. In this respect, the negative (positive) sign 
of the zero-field splitting parameter $D$ then corresponds to an easy-axis (easy-plane) uniaxial single-ion anisotropy. 

In the standard basis of spin states $|S_1^z, S_2^z \rangle$ ($S_1^z = \pm 1,0$ and $S_2^z = \pm 1,0$), the Hamiltonian (\ref{eq:ham}) can be defined through the following non-zero diagonal 
\begin{eqnarray}
\langle \pm 1, \pm 1 |\hat {\mathcal H}| \pm 1, \pm 1 \rangle &=& J + 2D \pm 2 H_z, \nonumber \\
\langle \pm 1, \mp 1 |\hat {\mathcal H}| \pm 1, \mp 1 \rangle &=& -J + 2D, \nonumber \\
\langle \pm 1, 0 |\hat {\mathcal H}| \pm 1, 0 \rangle &=& 
\langle 0, \pm 1 |\hat {\mathcal H}| 0, \pm 1 \rangle = D \pm H_z, 
\label{diag}	   
\end{eqnarray}
and non-diagonal matrix elements
\begin{eqnarray}
\langle \pm 1, 0 |\hat {\mathcal H}| \pm 1, \pm 1 \rangle &=& 
\langle \pm 1, 0 |\hat {\mathcal H}| \pm 1, \mp 1 \rangle = 
\langle 0, \pm 1 |\hat {\mathcal H}| \pm 1, \pm 1 \rangle = \nonumber \\
\langle 0, 0 |\hat {\mathcal H}| \pm 1, 0 \rangle &=& 
\langle 0, \pm 1 |\hat {\mathcal H}| \mp 1, \pm 1 \rangle = 
\langle 0, 0 |\hat {\mathcal H}| 0,  \pm 1 \rangle = \mbox{h.c.} = \frac{H_x}{\sqrt{2}}, 
\nonumber \\
\langle 1, -1 |\hat {\mathcal H}| \pm 1, \pm 1 \rangle &=& 
\langle -1, 1 |\hat {\mathcal H}| \pm 1, \pm 1 \rangle = \nonumber \\
\langle -1, 0 |\hat {\mathcal H}| 1, 0 \rangle &=& 
\langle 0, -1 |\hat {\mathcal H}| 0, 1 \rangle = \mbox{h.c.} = E,
\label{nondiag}	   
\end{eqnarray}
which are substantially simplified by the use of a new parameter $H_{\alpha} = g_{\alpha} \mu_{\mathrm{B}} B_{\alpha}$ ($\alpha=x,z$) that determines the effective magnetic field 
along two mutually orthogonal spatial directions. 

It is quite evident that an analytic diagonalization of the Hamiltonian defined through the matrix elements (\ref{diag}) and (\ref{nondiag}) is impossible, because the full Hamiltonian matrix cannot 
be reduced to a block-diagonal form with smaller-size matrices on its main diagonal. However, 
the nine-by-nine matrix can be rather easily diagonalized by making use of some exact numerical 
diagonalization method. In our case, the numerical diagonalization based on the Lapack subroutine 
DSYEV \cite{lapack}, which is suitable for real symmetric matrices, has been employed to calculate 
a complete set of eigenvalues and eigenvectors. Notice that all the other relevant quantities such as the partition function, Gibbs free energy, or magnetization, can be then straightforwardly computed 
with the help of standard thermodynamical-statistical relations.

\section{Numerical results and discussion}
\label{result}

Let us step forward to a discussion of the most interesting numerical results. Before doing this, 
it is worthy to notice that the effect of uniaxial single-ion anisotropy on the magnetization curves 
of the spin-1 Heisenberg dimer in the applied longitudinal magnetic field has been discussed in detail 
in our previous paper \cite{stre05}. The main objective of the present work is therefore to examine 
the magnetization curves of anisotropic spin-1 Heisenberg dimer in the transverse magnetic field, 
which should serve for the sake of a comparison with the relevant magnetization data acquired 
in the longitudinal field. 

\begin{figure}
\begin{center}
\includegraphics[width=14cm]{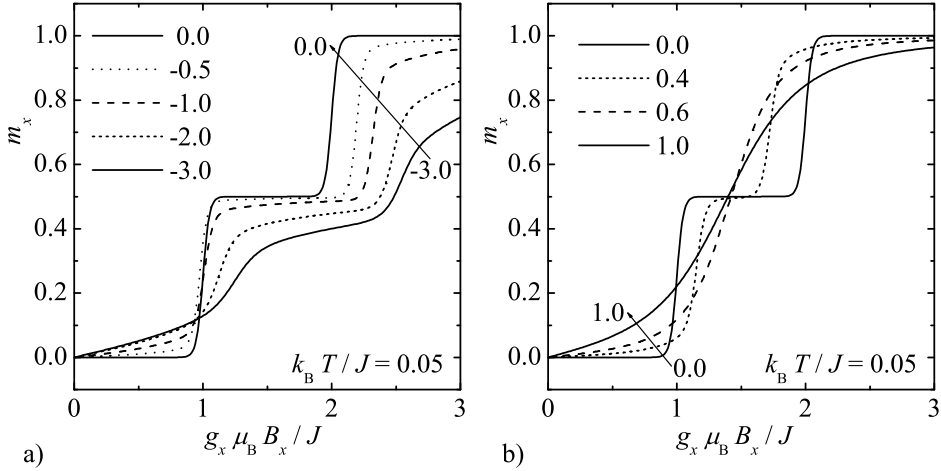} 
\end{center} 
\vspace{-9mm} 
\caption{Some typical low-temperature magnetization curves in the applied transverse magnetic 
field to be obtained for the uniaxial single-ion anisotropy of different strength. 
Fig.~\ref{fig1}a depicts magnetization curves for easy-axis anisotropy constants 
$D/J = 0.0$, $-0.5$, $-1.0$, $-2.0$, and $-3.0$, whereas Fig.~\ref{fig1}b shows magnetization 
curves for easy-plane anisotropy constants $D/J = 0.0$, $0.4$, $0.6$, and $1.0$ in ascending 
order along the direction of arrows.}
\label{fig1}
\end{figure}

Some typical low-temperature magnetization curves of the spin-1 Heisenberg dimer in the applied transverse magnetic field are depicted in Figs.~\ref{fig1}ab for several values of the parameter 
of the uniaxial single-ion anisotropy. Fig.~\ref{fig1}a shows the magnetization curves for the easy-axis single-ion anisotropy ($D<0$) that energetically favours the spin states $S_i^z = \pm 1$ 
with the smallest possible projection into the transverse field direction, while Fig.~\ref{fig1}b illustrates the effect of easy-plane single-ion anisotropy ($D>0$) that energetically favours the 
spin state $S_i^z = 0$ with the largest possible projection into the transverse field direction. 
In the former case, three magnetization plateaux (i.e. horizontal regions in the magnetization 
vs. magnetic field dependence) emerge at $m_x \equiv \langle S_i^{x} \rangle \approx 0.0$, $0.5$, 
and $1.0$ irrespective of the single-ion anisotropy strength, whereas in the latter case the intermediate magnetization plateau at a half of the saturation magnetization gradually diminishes 
by increasing $D/J$ until it completely vanishes from the magnetization curve above the threshold single-ion anisotropy $D_{t}/J = 0.60$. The breakdown of the intermediate magnetization plateau 
bears a close relation with the enhancement of the easy-plane single-ion anisotropy, which forces 
spins to lie in the $xy$-plane. Namely, it is quite reasonable to assume that the rising transverse field entails a rotation of spin precession axis towards the $x$-axis once the spin state 
$S_i^z = 0$ with the largest projection into the $xy$-plane is energetically preferred. 

Figs.~\ref{fig1}ab might also serve in evidence that the magnetization curves are gradually smeared
out upon strengthening the parameter of uniaxial single-ion anisotropy (i.e. by increasing 
the absolute value of the parameter $D/J$) even though one still might observe plateaux 
at particular magnetization values. It is quite plausible to argue that the gradual smoothing 
of magnetization curves occurs on behalf of a quantum entanglement, which is set up by the 
transverse magnetic field. Indeed, the transverse field triggers off a quantum entanglement 
between the spin states with unequal projections of the total quantum spin number and hence, 
the magnetization of antiferromagnetic state does not persist at the value $m_x = 0.0$, 
but it continuously increases with the transverse field even at zero temperature. Similarly, 
the magnetization of almost fully polarized and half-polarized state does not take precisely 
the values $m_x = 1.0$ and $0.5$, respectively, but it continuously varies with the transverse 
magnetic field for arbitrary but non-zero single-ion anisotropy. It is quite evident from Figs.~\ref{fig1}ab that the stronger is the single-ion anisotropy, the greater are 
the magnetization changes within three available plateau states with the magnetization 
$m_x \approx 0.0$, $0.5$, and $1.0$. It should be pointed out that the aforedescribed 
behaviour of the magnetization is in apparent contrast with what is observed in the applied 
longitudinal magnetic field. In a presence of the longitudinal field, the magnetization acquires 
at sufficiently low temperatures precisely one of three available values $m_z \equiv \langle S_i^{z} \rangle = 0.0$, $0.5$, $1.0$ and the stepwise magnetization curve with an abrupt change(s) 
of the magnetization at one or two transition fields must be consequently observed \cite{stre05}. 

\begin{figure}
\begin{center}
\includegraphics[width=8cm]{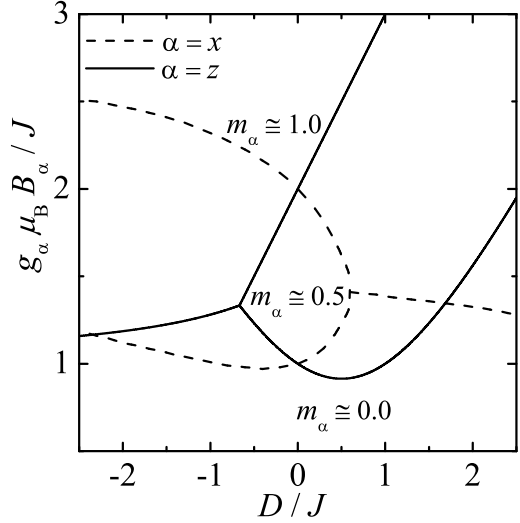} 
\end{center} 
\vspace{-9mm} 
\caption{The transition fields as a function of the uniaxial single-ion anisotropy in 
the magnetic field applied along the $x$- and $z$-axis, respectively. The parameter of biaxial 
single-ion anisotropy has been set to zero in this particular case.}
\label{fig2}
\end{figure}

Now, let us take a closer look at the difference between the magnetization curves in the 
applied longitudinal and transverse magnetic field, respectively. For comparison, Fig.~\ref{fig2} 
shows the transition fields as a function of the uniaxial single-ion anisotropy for two 
mutually orthogonal orientations of the applied magnetic field. It is worthy to mention that 
the displayed transition fields have been obtained as inflection points of the magnetization 
curves calculated at low enough temperature ($k_{\rm B} T/J = 0.05$). Note furthermore that 
the results for transition fields in the applied longitudinal field are thoroughly consistent 
with the ones calculated using the exact analytical diagonalization \cite{stre05}. It is quite 
obvious from Fig.~\ref{fig2} that the lines of transition fields divide the ground-state phase 
diagram into three different regions; the one appearing at sufficiently low magnetic fields 
corresponds to a weakly polarized state with almost perfect antiferromagnetic spin alignment 
$m_{\alpha} \approx 0.0$, the one appearing at sufficiently high fields corresponds to almost 
fully polarized state with the magnetization close to its saturation value $m_{\alpha} \approx 1.0$ 
and finally, the one appearing in the range of moderate fields corresponds to a half-polarized 
state with $m_{\alpha} \approx 0.5$. If the external field is oriented parallel with respect to 
a quantization axis (longitudinal field), then, it is even possible to derive exact analytical 
formulas for eigenfunctions that fully characterize three available ground states \cite{stre05} 
\begin{eqnarray}
|\Psi_{0,0} \rangle &=& \frac{1}{2} 
\left[A_{+} \left( |1,-1 \rangle + |-1,1 \rangle \right) - \sqrt{2} A_{-} |0,0 \rangle \right]; \label{gsa} \\
|\Psi_{1,1} \rangle &=& \frac{1}{\sqrt{2}} \left( |1,0 \rangle - |0,1 \rangle \right); \label{gsb}	\\
|\Psi_{2,2} \rangle &=& |1,1 \rangle. \label{gsc}	   
\end{eqnarray}
Notice that the probability amplitudes $A_{+}$ and $A_{-}$, which constitute the antiferromagnetic eigenstate (\ref{gsa}), are given by Eq.~(5) of Ref. \cite{stre05}. Altogether, it could be 
concluded that the magnetization curves in the applied longitudinal field distinguish 
the well-defined magnetization plateaux that are pertinent to the three available eigenstates (\ref{gsa})--(\ref{gsc}) with the magnetization per one site $m_{z} = 0.0$, $0.5$, and $1.0$, respectively (see for instance Fig.~3 in Ref. \cite{stre05}). Besides, it is also quite evident 
from Fig.~\ref{fig2} that the intermediate magnetization plateau at a half of the saturation magnetization gradually shrinks by decreasing the ratio $D/J$ until it completely vanishes from the magnetization curve below the threshold single-ion anisotropy $D_{\rm l}/J = -0.67$. The breakdown of intermediate magnetization plateau at sufficiently strong easy-axis single-ion anisotropies can 
readily be associated with a high energy cost of the spin state $S_i^z = 0$, which constitutes the eigenstate (\ref{gsb}) pertinent to the intermediate magnetization plateau. As a result, the 
single-ion anisotropy of two different types must be involved in order to cause a breakdown 
of the intermediate magnetization plateau in a presence of the longitudinal and transverse 
magnetic field, respectively. The disappearance of intermediate magnetization plateau in the 
applied longitudinal field takes place because of the easy-axis single-ion anisotropies 
$D< D_{\rm l}$, while the easy-plane single-ion anisotropies $D > D_{\rm t}$ are required 
to cause the breakdown of intermediate magnetization plateau in the transverse field direction.

For better orientation, the existence of the intermediate magnetization plateau under different 
spatial orientations of the applied magnetic field is summarized in Table 1 in relation to 
a strength of the uniaxial single-ion anisotropy. 
\begin{table}
\begin{center}
\begin{tabular}{|c|c|c|}
\hline  & \multicolumn{2}{|c|}{intermediate mg. plateau} \\	
\hline  $D/J$ & transverse (x) & longitudinal (z) \\	
\hline  (-$\infty$, -0.67) & yes & no \\	
\hline  (-0.67, 0.60) & yes & yes \\	
\hline  (0.60, $\infty$) & no & yes \\ \hline
\end{tabular}
\caption{The appearance of the intermediate magnetization plateau in dependence on a strength 
of the uniaxial single-ion anisotropy $D/J$ under two mutually orthogonal spatial orientations 
of the applied magnetic field.}
\end{center}
\end{table}
The intermediate magnetization plateau can be accordingly found in both conspicuous spatial 
directions only if the uniaxial single-ion anisotropy $D$ is small enough compared to 
the exchange interaction $J$. Otherwise, the intermediate plateau is present either in the 
applied longitudinal field and is simultaneously absent in the transverse field, or vice versa. 
Thus, one might conclude that the sufficiently strong uniaxial single-ion anisotropy (no matter 
whether of easy-axis or easy-plane type) entails a rather striking magnetization process, 
which is characterized by two qualitatively different magnetization curves (to be obtained 
for two mutually orthogonal orientations of the applied magnetic field) with and 
without the intermediate magnetization plateau. 

\begin{figure}
\begin{center}
\includegraphics[width=14cm]{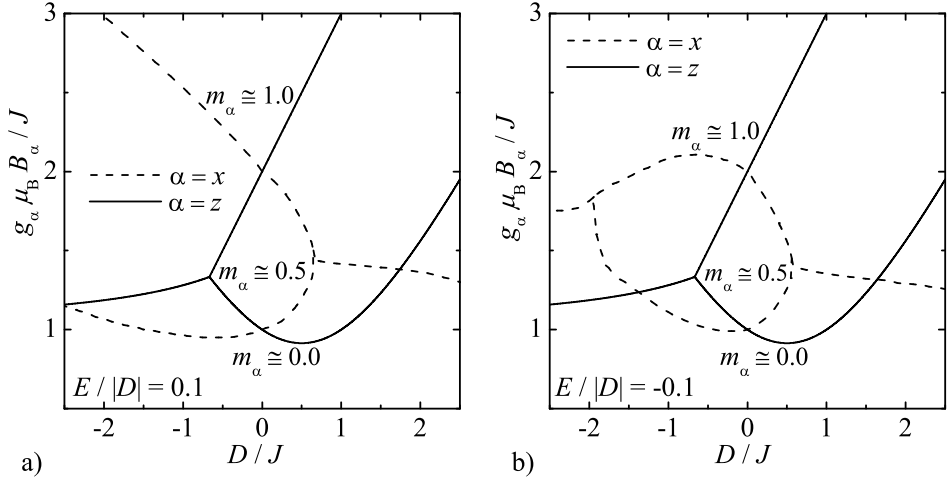} 
\end{center} 
\vspace{-9mm} 
\caption{The transition fields as a function of the uniaxial single-ion anisotropy in the magnetic field applied along the $x$- and $z$-axis, respectively. The constant ratio between the uniaxial 
and biaxial single-ion anisotropy parameters has been set to: a) $E/|D| = 0.1$; b) $E/|D| = -0.1$.}
\label{fig3}
\end{figure}

The situation becomes even much more intriguing by turning on the biaxial single-ion anisotropy.
To illustrate the case, we depict in Figs.~\ref{fig3}ab transition fields as a function of the uniaxial single-ion anisotropy for two different spatial orientations of the applied magnetic field and 
two different values of the ratio $E/|D| = 0.1$ and $-0.1$. It is worthwhile to remark that the
negative (positive) value of the anisotropy parameter $E$ facilitates (hampers) magnetization 
along the $x$-axis in the otherwise magnetically isotropic $xy$-plane. Apparently,
the biaxial single-ion anisotropy has just minor effect upon the transition fields in the 
applied longitudinal field (the relevant data cannot be distinguished within the displayed 
scale from the ones calculated for the zero biaxial anisotropy) notwithstanding the character 
(sign) of the anisotropy parameter $E$. It turns out, however, that the biaxial 
single-ion anisotropy may basically change the transition fields in the applied transverse 
magnetic field. If the $x$-axis is the hard magnetization axis within the $xy$-plane, 
i.e. if $E>0$, then, the relevant transition fields are merely slightly shifted from the 
ones calculated for $E=0$ and one still observes qualitatively the same dependences. 
As a matter of fact, the only sizeable change of transition fields can be detected in the 
field-induced transition from the intermediate plateau state to the saturated paramagnetic 
state, which shows steeper variation of the transition fields in the range of negative 
uniaxial single-ion anisotropies (cf. Fig.~\ref{fig3}a with Fig.~\ref{fig2}). By contrast, 
the transition field vs. single-ion anisotropy dependence changes qualitatively rather than quantitatively by considering the biaxial single-ion anisotropy $E<0$, which alters the $x$-axis 
into the easy magnetization axis within the $xy$-plane. In such a case, the emergence of 
intermediate plateau is restricted just to a certain interval of the single-ion anisotropies 
($-2.0 \lesssim D/J \lesssim 0.5$), whose endpoints show a weak dependence on the parameter $E$. 
The most interesting finding to observe here is a peculiar breakdown of the intermediate
magnetization plateau, which results from the non-zero biaxial single-ion anisotropy on assumption 
that there is a strong enough easy-axis anisotropy $D/J \lesssim -2.0$ and one applies 
the transverse magnetic field along the easier magnetization axis within the $xy$-plane. 

\section{Theory vs. experiment}           

\begin{figure}
\begin{center}
\hspace*{7mm}
\includegraphics[width=14.5cm]{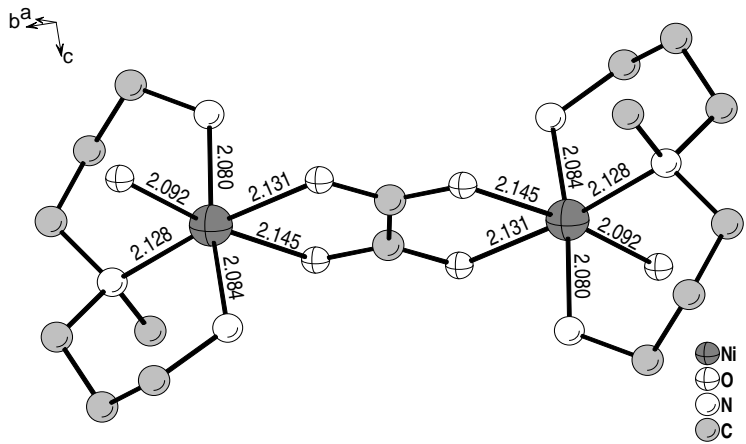} 
\end{center} 
\vspace{-13.8cm} 
\caption{The homodinuclear complex cation [Ni$_2$(Medpt)$_2$($\mu$-ox)(H$_2$O)$_2$]$^{2+}$
constituting the dinuclear core of NAOC coordination compound. This figure was drawn using the DIAMOND programme \cite{diamond}.}
\label{fig4}
\end{figure}

At this stage, let us compare the obtained theoretical results with the relevant experimental 
magnetization data of NAOC compound.\footnote{High-field magnetization measurements performed 
on NAOC compound have been originally reported by some of the present authors in the earlier publications \cite{Narumi}--\cite{ESR} to which the interested reader is referred to for
closer experimental details.} Before proceeding to the relevant comparison, however, 
few remarks should be made on a crystal structure of this coordination compound to enable 
a deeper insight into its magneto-structural correlations. The single-crystal sample of NAOC 
is essentially an assembly of dinuclear complex cations, which are held together merely 
by virtue of perchlorate counter anions, hydrogen bonds and van der Waals contacts \cite{EVR}. 
Fig.~\ref{fig4} shows a schematic view on the discrete dinuclear unit and it also specifies all 
bond lengths incident to paramagnetic nickel centres. As one can see, both octahedrally coordinated 
nickel centres are linked through the bis-chelating oxalato group, whereas the rest of their coordination sphere is completed by the blocking tridentate amine (Medpt) and one ligating 
water molecule. It can be also clearly seen from Fig.~\ref{fig4} that the complex cation of NAOC 
represents a centrosymmetric unit with an inversion center at the midpoint of the bridging 
oxalate group. This has far-reaching consequences on possible sources of the magnetic anisotropy. 
The antisymmetric Dzialoshinskii-Moriya interaction \cite{DM} must be entirely disregarded 
due to a presence of the inversion center and moreover, our previous study has revealed 
the totally negligible exchange anisotropy (i.e. the anisotropy in a symmetric pseudodipolar interaction) as well \cite{stre05}. The most crucial contribution to the overall magnetic 
anisotropy should be therefore related to the single-ion anisotropy, which is closely 
connected to the crystal field of ligands surrounding the paramagnetic nickel centres. 
Altogether, the structural data 
listed in Fig.~\ref{fig4} might indicate a relatively strong uniaxial single-ion anisotropy 
due to a rather high difference between bond lengths to the axial and equatorial ligands, 
respectively. Besides, somewhat smaller biaxial single-ion anisotropy might be also expected 
on account of diverse bond distances in the equatorial plane. Actually, the bond distance 
to the coordinated water molecule is slightly shorter than the bond distances to other three 
equatorial ligands.

\begin{figure}
\begin{center}
\includegraphics[width=14cm]{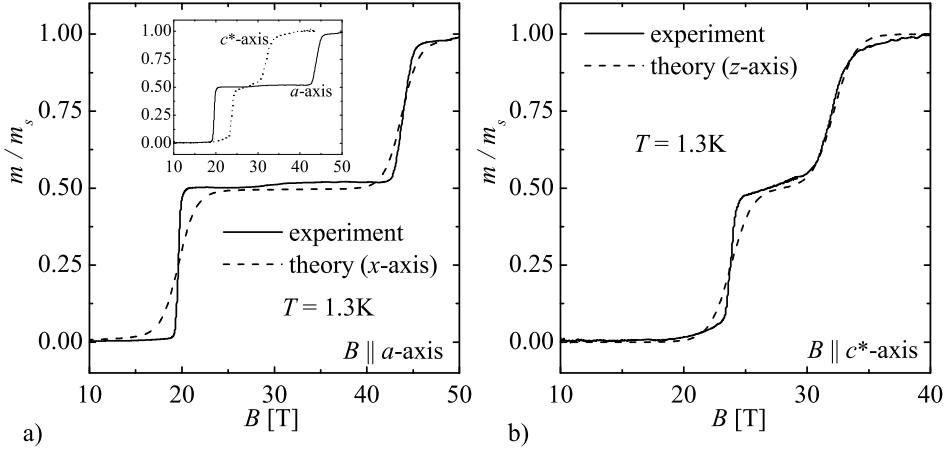} 
\end{center} 
\vspace{-9mm} 
\caption{The high-field magnetization curves measured at sufficiently low temperature ($T$=1.3K)
in the magnetic field applied along the crystallographic $a$- and $c^{*}$-axis, respectively.
The best simultaneous fit of both these magnetization curves has been achieved through this
unique set of fitting parameters: $J/k_{\rm B} = 30.7$K, $D/k_{\rm B} = -12.5$K, and $g = 2.26$.
Both magnetization curves are displayed within the same scale in the insert of Fig.~\ref{fig5}a 
in order to show a substantial anisotropy in the magnetization process.}
\label{fig5}
\end{figure}

\begin{figure}
\begin{center}
\includegraphics[width=14cm]{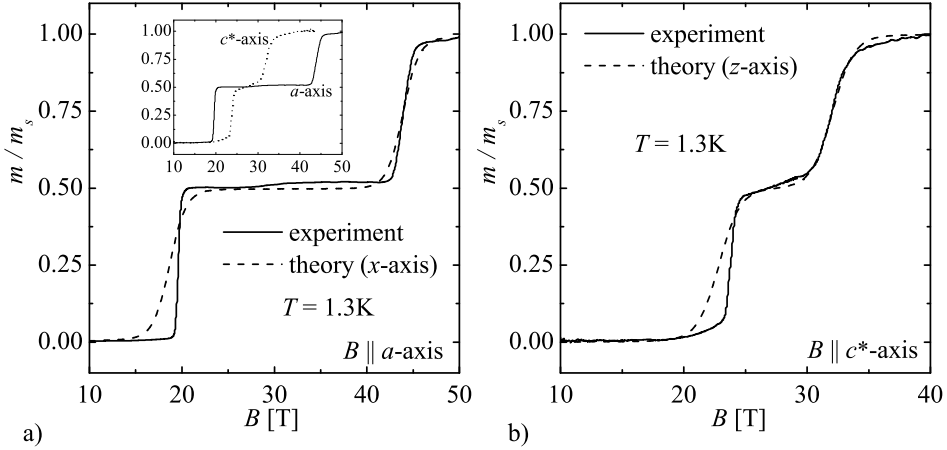} 
\end{center} 
\vspace{-9mm} 
\caption{The high-field magnetization curves measured at sufficiently low temperature 
($T$=1.3K) in the magnetic field applied along the crystallographic $a$- and $c^{*}$-axis, 
respectively. The best simultaneous fit of both these magnetization curves has been achieved 
through this unique set of fitting parameters: $J/k_{\rm B} = 30.2$K, $D/k_{\rm B} = -11.0$K, 
$E/k_{\rm B} = 1.5$K, and $g = 2.20$.}
\label{fig6}
\end{figure}

The high-field magnetization data measured in a pulsed magnetic field, which is oriented either 
along the crystallographic $a$- or $c^{\star}$-axis ($a \perp c^{\star}$), are plotted in Figs.~\ref{fig5} and \ref{fig6} together with the corresponding theoretical predictions for 
the longitudinal and transverse magnetization. Fig.~\ref{fig5} shows the best simultaneous fit 
of both experimental magnetization curves by considering only the parameters $J$, $D$, and $g$ pertinent to a strength of the exchange interaction, uniaxial single-ion anisotropy, and 
gyromagnetic ratio, respectively, as the adjustable fitting parameters and neglecting the 
biaxial single-ion anisotropy as a higher-order anisotropy term. Even under this restriction, 
the results obtained from the exact numerical diagonalization directly remove ambiguity in 
determining a strength of the uniaxial single-ion anisotropy, which otherwise occurs in an attempt 
to fit both magnetization curves through a single expression for the longitudinal magnetization 
known from the exact analytical diagonalization (see for details Ref.~\cite{stre05}). According 
to this plot, the crystallographic $a$-, $b$- and $c^{\star}$-axes definitely turn out to be the 
$x$-, $y$- and $z$-axes of the effective spin Hamiltonian (\ref{eq:ham}) and the negative sign 
of the parameter $D$ is also consistent with the experimental observation that the $c^{\star}$-axis 
is the easy magnetization axis \cite{ESR}. The best result of a more comprehensive fitting procedure, which includes the parameter $E$ to the set of adjustable fitting parameters, is depicted in Fig.~\ref{fig6}. 
Evidently, the inclusion of the parameter of biaxial single-ion anisotropy merely causes a small 
reduction of the most dominant interaction parameters $J$, $D$, $g$ (cf. the fitting data sets  
listed in the last two figure captions) and beside this, the fitting data set presented in 
Fig.~\ref{fig6} is in an excellent agreement with the one proposed by the analysis of high-field 
ESR data \cite{ESR}. Both these facts clearly demonstrate adequate reliability and plausibility 
of the data sets to be obtained by fitting. When comparing the accuracy of both fitting sets 
in reproducing the experimental data, the latter data set fits magnetization curves more 
precisely in the region of intermediate magnetization plateau and around the second transition 
field, while the former one gives a more adequate description in a vicinity of the first 
transition field that is slightly underestimated in the latter fitting set. It is worthwhile 
to remark, however, that the magnetization curves measured in a pulsed magnetic field are 
not absolutely isothermal ones on behalf of a magnetocaloric effect that cools down spin 
system in a field ascending process close to the transition fields. This fact might be 
regarded as a possible reason for slightly sharper dependence of the experimental magnetization 
curves to be observed especially near the first transition field. Finally, it is quite 
plausible to argue that the actual fitting set should be capable of rendering the overall 
angular dependence of transition fields. Namely, this kind of fitting would represent 
perhaps the most efficient way how to obtain indisputable fitting set, because there 
is still a certain danger of overinterpretation when attempting to fit two magnetization 
curves through four adjustable fitting parameters. The fitting of the overall angular 
dependence of transition fields is nevertheless a rather time-consuming numerical problem, 
which is left as a challenging task for our future work. 

\section{Conclusion}
\label{conclusion}

The present article provides a deeper insight into the magnetization process of the spin-1 
Heisenberg dimer model with either uniaxial or biaxial single-ion anisotropy. The main motivation 
to investigate this rather simple quantum spin model bears a close relationship with the fact 
that it serves as a versatile model for a large number of existing dinuclear coordination compounds containing the dinickel core. Within the framework of the exact numerical diagonalization method, 
we have constructed for the homodinuclear nickel(II) complexes essentially exact ground-state phase diagrams in the form of transition field vs. single-ion anisotropy dependence (Figs.~\ref{fig2} 
and \ref{fig3}). The most important finding to emerge from the present study is closely associated 
with a theoretical prediction concerning the possible breakdown of intermediate magnetization plateau. If the uniaxial single-ion anisotropy is strong enough with respect to the exchange interaction, 
we have found convincing evidence that a presence of the intermediate magnetization plateau 
in the applied longitudinal field demands its absence in the applied transverse field 
and vice versa. Owing to this fact, it could be concluded that a sufficiently strong uniaxial single-ion anisotropy (regardless of whether easy-axis or easy-plane type) is responsible 
for a striking magnetization process, which is characterized by two qualitatively different magnetization curves (to be obtained for two mutually orthogonal directions of the applied 
magnetic field) with and without the intermediate magnetization plateau.  

The obtained theoretical results were also compared with the high-field magnetization measurements 
on the homodinuclear nickel(II) coordination compound NAOC, which is being regarded as a typical experimental realization of the spin-1 Heisenberg dimer model. The best simultaneous fit of magnetization data, which were measured in the magnetic field applied along two mutually orthogonal crystallographic $a$- and $c^{\star}$-axes, was attained for the fitting set $J/k_{\rm B} = 30.2$K, $D/k_{\rm B} = -11.0$K, $E/k_{\rm B} = 1.5$K, $g = 2.2$ that is consistent with the one reported 
on previously by the analysis of high-field ESR data \cite{ESR}. The actual value of the relative 
ratio $D/J \cong -0.36$ implies for NAOC compound a relatively strong uniaxial single-ion anisotropy 
of the easy-axis type, which is, on the other hand, too small to cause the breakdown of intermediate magnetization plateau theoretically predicted (the threshold value is $D_{\rm l}/J = -0.67$). 
From this point of view, high-field magnetization measurements on another single-crystal samples prepared from an immense reservoir of homodinuclear nickel(II) complexes \cite{stre05} would be desirable in order to provide an experimental confirmation of this interesting quantum phenomenon. 
Even though a design of molecule-based magnetic materials with a tunable strength of the exchange interaction and single-ion anisotropy is far from being a routine target at present, the homodinuclear nickel(II) complexes with a less rigid bridging group (i.e. with a possibly weaker exchange interaction) and simultaneously a higher distortion of coordination octahedron (i.e. with 
a possibly stronger single-ion anisotropy) might be considered as suitable candidates for 
displaying such an interesting quantum phenomenon. 

\ack{J. Stre\v{c}ka would like to thank Japan Society for the Promotion of Science 
for the award of postdoctoral fellowship (ID No. PE07031) under which part of this work 
was carried out. This work was partially supported also by the Slovak Research and 
Development Agency under the contracts LPP-0107-06 and APVT 20-005204.}

\end{document}